\documentclass[aps, prx, reprint, longbibliography, nofootinbib, floatfix,superscriptaddress,]{revtex4-2}

\usepackage{ragged2e}
\usepackage{slashed}
\usepackage{verbatim}
\usepackage[T1]{fontenc}
\usepackage[utf8]{inputenc}
\usepackage[colorlinks]{hyperref}
\usepackage{mathbbol}
\usepackage[dvipsnames]{xcolor}
\usepackage{graphicx}
\usepackage{orcidlink}
\usepackage{euscript}
\usepackage[normalem]{ulem}
\usepackage{svg}
\usepackage[english]{babel}
\usepackage{lipsum}
\usepackage{physics}
\usepackage{dcolumn}
\usepackage{tensor}
\usepackage{comment}
\usepackage{graphicx,color,overpic,mathtools}
\usepackage{amsthm,amsmath,amssymb,hyperref,mathrsfs}
\usepackage{braket,bm,bbm,setspace}
\usepackage{cancel}
\usepackage{float}
\usepackage{galdrastafir}

\hypersetup{
    colorlinks=false,
    pdfborder={0 0 0},
}
\usepackage{xargs}
\usepackage{tikz}

 \hypersetup{
    colorlinks=true,
    linkcolor=oxblood, 
     citecolor=oxblood, 
    urlcolor=oxblood   
 }

\usepackage{staves}

\newcommand{\marksize}{\fontsize{4}{5}\selectfont}

\makeatletter
\def\@fnsymbol#1{%
  \ifcase#1\or
    \marksize\staveI\or
    \hulinhjalmur[size=1.1em]\or
    \marksize\staveXXIV\or
    \marksize\staveV\or
       
    \marksize\staveII\or
    \marksize\staveVI\or
    \marksize\staveVII\or
    \marksize\staveVIII\or
    
    \marksize\staveIII\or
    
    \marksize\staveIV\or
    \marksize\staveIX\else
    
    \@ctrerr
  \fi
}
\makeatother

\usetikzlibrary{arrows.meta,calc,angles,quotes,shapes.geometric,decorations.markings}

\definecolor{mypurple}{RGB}{130, 0, 130} 
\definecolor{oxblood}{rgb}{0.5333, 0.0314, 0.0314}
\definecolor{orange}{rgb}{1,0.5,0}

\begin{document}

\title{Strong Field Gravitational Lensing in Full Wave Optics}

\author{Jo\~ao S. Santos\textsuperscript{\protect\orcidlink{0000-0003-1921-8758}}}
\email{joao.sieiro.santos@tecnico.ulisboa.pt}
\affiliation{CENTRA, Departamento de Física, Instituto Superior Técnico -- IST,
Universidade de Lisboa -- UL, Avenida Rovisco Pais 1, 1049-001 Lisboa, Portugal}
\affiliation{Center of Gravity, Niels Bohr Institute, Blegdamsvej 17, 2100 Copenhagen, Denmark}

\author{Luka Vujeva\,\orcidlink{0000-0001-7697-8361}}
\email{luka.vujeva@nbi.ku.dk}
\affiliation{Center of Gravity, Niels Bohr Institute, Blegdamsvej 17, 2100 Copenhagen, Denmark}

\begin{abstract}

Compact binaries evolving in the vicinity of massive black holes may be commonplace in the Universe, and their gravitational waves carry signatures of their environment. We study the strong lensing of these signals by the massive
companion, a configuration that violates the standard assumptions of the lensing literature: geometric optics and thin lenses. Using a perturbative framework, we compute the full wave optics amplification factors for face-on and edge-on binaries, including spinning lenses. In the edge-on configuration, peak amplification does \textit{not} occur at perfect source--lens--observer alignment, even for a non-spinning lens. This is a consequence of the anisotropic, coherent quadrupolar emission of the stellar-mass binary breaking the axisymmetry of the system. Restricting to the equatorial plane near the optical axis, we then find that the frequency dependence of the amplification factor is universal across every configuration studied, persisting deep into the wave optics regime. Comparing against a strong-field ray tracing calculation, we find agreement at the $\sim10\%$ level for a face-on binary lensed by a Schwarzschild black hole. In the less symmetric edge-on case only the angular peak shift and fringe spacing is reproduced to this accuracy. For a Kerr lens, the face-on amplification factor is described by its Schwarzschild counterpart to the same $\sim 10 \%$ accuracy once the effect of frame dragging is corrected for. A point mass lens, by contrast, fails to reproduce the features of the system, which may have strong implications for forecasts that adopt it for lensing by massive black holes in the strong field.
\end{abstract}

\maketitle

%
\section{Introduction}
%
We have entered the second decade of gravitational wave (GW) astronomy. Yet, in over 300 compact binary coalescences~\cite{LIGOScientific:2026wfs}, we have yet to find an event that shows compelling evidence of deviating from the isolated vacuum, general relativistic, 2-body merger hypothesis~\cite{LIGOScientific:2026qni, LIGOScientific:2026oim, LIGOScientific:2025rid, LIGOScientific:2025cwb, LIGOScientific:2026pwx, Roy:2026mco}. The increased sensitivity and frequency coverage  of future detectors~\cite{LISA:2024hlh,Kawamura:2020pcg,Gupta:2023lga, ET:2025xjr,TianQin:2015yph,Evans:2021gyd} will eventually capture deviations from this hypothesis. These may be corrections to general relativity, signatures of the binary's astrophysical environments, GW lensing, or any effect lurking in the next decimal place. 

An exciting prospect is the detection of a stellar mass binary black hole (BH) merger in the vicinity of a massive black hole (MBH). These triple systems, often referred to as binary extreme mass ratio inspirals~\cite{Chen:2018axp,Han:2018hby,Cardoso:2021vjq,Yin:2024nyz,Santos:2025ass}, may be formed in the accretion disks of active galactic nuclei powered by MBHs~\cite{Bartos:2016dgn,Stone:2016wzz,Dittmann:2023sha, Peng:2024wqf,Vaccaro:2025dzl,Tagawa:2019osr,Samsing:2020tda, Hoang:2017fvh}. Compact objects are captured by the disk, driven by torques into traps at distances of $\mathcal{O} (10-1000)$ gravitational radii~\cite{Bellovary:2015ifg, Peng:2021vzr, Yunes:2011ws, Pan:2021ksp, Derdzinski:2022ltb}, where they meet and form binaries. This dynamical formation channel for binary BHs would also be a compelling mechanism to produce hierarchical mergers and the observed high mass BH which are hard to explain under the standard stellar binary evolution pictures~\cite{Gerosa:2021mno, Gerosa:2017kvu,LIGOScientific:2025brd, LIGOScientific:2025rsn, LIGOScientific:2026ctl}. 

Mergers near MBHs offer a compelling opportunity to study strong field gravity effects, namely the scattering of gravitational waves off BHs. In recent years, this topic has garnered a great deal of interest, mostly from two communities, each tackling the problem in complementary ways: BH perturbation theory and weak field lensing. The perturbative approach is particularly suited to the low frequency regime of the problem both analytically~\cite{Dolan:2007ut, Dolan:2008kf, Pijnenburg:2024btj, Kubota:2024zkv}, and numerically~\cite{Nambu:2012wa, Folacci:2019vtt, Chan:2025wgz, Li:2025gou, Saketh:2025cwf}. The weak field approach uses the arrival time difference of light rays in the vicinity of the traditional lensing images to find the frequency dependent lensing effects~\cite{Oancea:2022szu,Oancea:2023hgu,DOrazio:2019fbq,Ubach:2025oxr, Martens:2026mku,Huang:2026hbb}. In other words, one approach starts in the low frequency, strong field regime, and the other starts in the infinite frequency, weak field regime. This makes direct, meaningful comparisons particularly challenging. On the other hand, recent works have aimed to quantify the rate at which such lensing from binaries in the accretion disks of active galactic nuclei can be seen~\cite{Leong:2024nnx}. However, they rely on Newtonian approximations of the lensing potential, which may not be valid at the radii relevant to astrophysical binaries near MBHs. This serves as further motivation to pursue a more robust treatment of wave optics (WO) effects in such regimes.

In this work, we aim to tackle both these problems by following two distinct approaches to the BH lensing problem. The first uses uses BH perturbation theory~\cite{Santos:2025ass} and is especially suited to low frequency, strong field, scattering; the second is built using null geodesics in the infinite frequency regime, more reminiscent of weak field approaches to WO~\cite{Takahashi:2003ix,Ezquiaga:2025gkd}. Besides matching the two regimes, we perform a characterization of some of the phenomenology of this extreme setup, including the effects of the MBH spin and of the small binary spin orientation relative to the lens and observer.

Finally, we quantify the validity of employing weak field approximations deep into the strong field regions where migration traps populated by binary systems may exist. We also provide corrections to the weak field methodology, which can be used to mitigate the errors associated with naively using these in the strong field regime. While this approach cannot reproduce the full WO structure, it can serve as a cost effective improvement to current search efforts.

The paper is organized as follows. Section~\ref{sec:framework} outline the two distinct methods in which we approach the scattering problem. First, we review the basics of the relativistic scattering approach in Sec.~\ref{sec:b_EMRI} (for a more detailed description of this model, see Ref.~\cite{Santos:2025ass}). Next, we describe the complementary approach built using ray tracing techniques (more details can be found in Ref.~\cite{Vujeva:2026rqh}). Sec.~\ref{sec:results} compares our two independent approaches in their domains of validity, as well as comparing to common Newtonian treatments used in GW lensing literature. Finally, Sec.~\ref{sec:conclusions} provides a summary of the findings of this work as well as future prospects for GW lensing in the strong field.

\section{Framework} \label{sec:framework}
%
\subsection{The geometric setup} \label{sec:setup}
\begin{figure}[ht!]
\usetikzlibrary{arrows.meta,calc,angles,quotes,shapes.geometric}

\definecolor{myred}{HTML}{C40000}
\definecolor{myblue}{HTML}{0053A1}
\definecolor{mygray}{HTML}{8A8A8A}

\begin{tikzpicture}[
    >=Latex,
    font=\normalsize,
    axis/.style={->, thin},
    ray/.style={thin, mygray},
    guide/.style={dashed, thin, mygray},
    anglemark/.style={->, thin, mygray},
    spin/.style={->, line width=0.9pt, myblue},
    sbi/.style={->, thin},
    reddir/.style={->, line width=0.9pt, myred},
    dot/.style={circle, fill=black, inner sep=0pt, minimum size=6pt},
    source/.style={circle, draw=black, fill=white, line width=0.8pt,
                   inner sep=0pt, minimum size=7pt},
    lens/.style={circle, fill=black, inner sep=0pt, minimum size=48pt}
]

\coordinate (O)   at (0,0);
\coordinate (Z)   at (0,2.0);
\coordinate (X)   at (3.8,0);
\coordinate (Y)   at (2.3,1.55);

\coordinate (Obs) at (-2.3,1.65);
\coordinate (Proj) at (-2.3,0.7);

\coordinate (Src) at (2.5,0,0);

\draw[axis] (O) -- (Z) node[above] {$z$};
\draw[axis] (O) -- (X) node[right] {$x$};
\draw[axis] (O) -- (Y) node[above right] {$y$};

\coordinate (Eye) at ($(Obs)+(0.06,0.10)$);

\begin{scope}[shift={($(Obs)+(-.3,0.2)$)}]
    \draw[fill=white] (0,0.10) circle (0.10);

    \draw (0,0.00) -- (0,-0.28);

    \draw (-0.16,-0.10) -- (0,-0.02) -- (0.16,-0.10);

    \draw (0,-0.28) -- (-0.14,-0.46);
    \draw (0,-0.28) -- (0.14,-0.46);
\end{scope}

\node[above=7pt] at ($(Obs)+(0,0.22)$) {\footnotesize Obs.$(r_{\rm O},\theta_{\rm O},\phi_{\rm O})$};

\draw[ray] (Eye) -- (O);
\draw[guide] (Eye) -- (Proj) -- (O);
\node[lens] at (O) {};
\node[text=white] at (O) {$M,a$};

\draw[anglemark] ($(O)+(0,0.95)$)
    arc[start angle=90, end angle=145, radius=0.95];
\node[mygray] at (-0.55,1.05) {$\theta_0$};

\begin{scope}[shift={(O)}, x={(1.2cm,0cm)}, y={(0.1cm,0.40cm)}]
    \draw[anglemark] (1.0,0)
        arc[start angle=0, end angle=-220, radius=1.1];
    \node[mygray] at (-1.4,-1) {$2\pi-\phi_0$};
\end{scope}
(2.3,1.55)
\node[source] at (Src) {};

\draw[sbi] ($(Src)+(0,0.20)$)
    .. controls +(0.10,0.70) and +(-0.50,-0.10) ..
    ($(Src)+(0.80,0.80)$);

\node[left=2pt] at ($(Src)+(0.0,0.65)$) {\footnotesize SB I};

\draw[sbi] ($(Src)+(0,-0.20)$)
    .. controls +(0.10,-0.70) and +(-0.50,0.10) ..
    ($(Src)+(0.80,-0.80)$);

\node[left=1pt] at ($(Src)+(0.0,-0.75)$) {\footnotesize SB II};

\coordinate (UC) at (4.0,1.0);

\begin{scope}[shift={(UC)}, x={(0.14cm,0.10cm)}, y={(0cm,0.55cm)}]

    \draw[thick] (0:.5) arc[start angle=0, end angle=360, radius=.5];

    \draw[thick] (120:.5) arc[start angle=120, end angle=300, radius=.5];
    \draw[thick] (300:.5) arc[start angle=300, end angle=480, radius=.5];

    \node[dot] at (90:.5) {};
    \node[dot] at (270:.5) {};

\end{scope}

\draw[reddir] (UC) -- ($(UC)+(-0.6,0)$);

\coordinate (LC) at (4.,-1);

\begin{scope}[shift={(LC)}, x={(0.65cm,0cm)}, y={(0.04cm,0.18cm)}]

    \draw[thick] (0:.5) arc[start angle=0, end angle=360, radius=.5];

    \draw[thick] (140:.5) arc[start angle=140, end angle=300, radius=.5];
    \draw[thick] (300:.5) arc[start angle=300, end angle=500, radius=.5];

    \node[dot] at (200:.5) {};
    \node[dot] at (20:.5) {};

\end{scope}

\draw[reddir] (LC) -- ($(LC)+(0,0.6)$);
\end{tikzpicture}
\caption{\justifying 
The geometrical setup that we consider here.
The primary is a massive Kerr BH with mass $M$ and spin $a$ pointing in the $\hat z$ direction of the diagram. Consider $\{t,r,\theta,\phi\}$ Boyer-Lindquist coordinates centered at the massive BH. 
A stellar-mass binary (SB) is placed, stationary, on the $x$ axis, with coordinates $(r_{\rm S}, \pi/2, 0)$, and with its spin $\vec S$ (red arrow) either pointing in the $-\hat x$ (SB I) or in the $\hat z$ direction (SB II). The observer is at coordinate position $(r_{\rm O},\theta_{\rm O},\phi_{\rm O})$, with $r_{\rm O}\gg M$, and the angles illustrated in gray. The SB emits gravitational waves which are scattered by the massive BH and before reaching the observer. 
}
\label{fig:binary_geometry}
\end{figure}
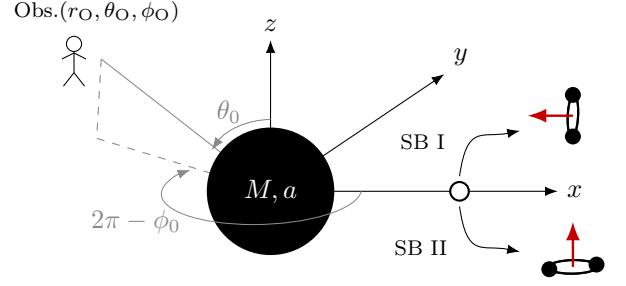

%
%
%
In this paper we study the setup depicted in Fig.~\ref{fig:binary_geometry}, composed of an equal mass binary (hereafter the stellar-mass binary, SB), in the spacetime of a Kerr MBH with mass $M$ and spin parameter $a$. In Boyer-Lindquist coordinates $\{t,r,\theta,\phi\}$, the Kerr metric is written as
\begin{eqnarray}
    ds^2 &=& - \left(1-\frac{2Mr}{\Sigma}\right) dt^2 - \frac{4 M a r \sin^2 \theta}{\Sigma} dt d\phi+\frac{\Sigma}{\Delta} dr^2  \nonumber \\ 
& +& \Sigma d\theta^2 + (r^2 +a^2 
+ \frac{2 M r a^2 }{\Sigma} \sin^2 \theta) \sin^2 \theta \; d\phi^2 , \label{eq:kerr_metric}
\end{eqnarray}
where
\begin{equation}
\Sigma =r^2 + a^2 \cos^2 \theta \, , \quad \Delta = r^2 -2 M r + a^2
\end{equation}
and $0\leq a  \leq M$. The inner and outer horizons are located at the roots of $\Delta$,
\begin{equation}
r_\pm = M \pm \sqrt{M^2 -a^2}\, .
\end{equation}
The center of mass (CoM) of the SB is stationary\footnote{The model developed in Ref.~\cite{Santos:2025ass} allows for CoM to follow a circular equatorial geodesic. We also take the SB to be on the $\phi=0$ line. Here we choose to focus on the simpler stationary case because this makes the signal monochromatic, simplifying the analysis and interpretation of results.}, in the equatorial plane $\theta=\pi/2$, and at $r=r_{\rm S} > 2 M$,  but fully in the strong field region. Each component of the SB has mass $M_{\rm SB}/2$, and these components are in circular motion around the CoM with separation $d_{\rm SB}$. We focus on scenarios with a sharp hierarchy of scales, $M_{\rm SB} \ll d_{\rm SB} \ll M <r_{\rm S}$, as this is an assumption of the framework developed in~\cite{Santos:2025ass}.
We will focus in two geometric configurations for the system: SB I where the spin of the SB, $\vec S$, points in the direction of the MBH, and SB II where it points along the $\hat{z}$ direction (see Fig.~\ref{fig:binary_geometry}). The observer is at coordinates $(r_{\rm O},\theta_{\rm O},\varphi_{\rm O})$, where we are interested in $r_{\rm O} \gg M,r_{\rm S}$, and vary its angular position relative to the lens. 
\begin{table}[]
    \centering
    \begingroup
\setlength{\tabcolsep}{6pt} 
\renewcommand{\arraystretch}{1.2} 
\begin{tabular}{l|c|c}
    \textbf{Quantity} & \textbf{Symbol} & \textbf{Fiducial} \\
    \hline
        MBH mass & $M$ & 1 \\ 
        MBH spin & $a$ & --- \\
        MBH coordinates & $\{t,r,\theta,\phi\}$ & --- \\
        Source radius & $r_{\rm S}$ & 10 \\
        Source component mass & $M_{\rm SB}/2$ & $10^{-4}$ \\
        Source separation & $d_{\rm SB}$ & $0.05$ \\
        Source orbital frequency & $ {\rm \Omega_{SB}} $ & 1.13 \\
        Source coordinates & $\{T, R, \vartheta_{\rm S}, \varphi_{\rm S}\}$ & ---\\
        Source angular position  & $ (\theta_{\rm S}, \phi_{\rm S}) $ & $(\pi/2,0)$ \\ 
        GW frequency & $\omega$ & 2.26 \\ 
        Amplification factor  & $F$ & --- \\
        Observer radius & $r_{\rm O}$ & $\infty$ \\ 
        Observer angular position  & $ (\theta_{\rm O}, \phi_{\rm O}) $ & --- \\ 
        \hline
    \end{tabular}
\endgroup
    \caption{\justifying All symbols for the system parameters in this work, including their values for the fiducial simulation. These correspond, for example, to a SB composed of two $100 M_\odot$ BHs near a $10^6 M_\odot$ MBH. }
    \label{tab:parameters_fiducial}
\end{table}
%
%
\subsection{The full wave optics model} \label{sec:b_EMRI}
%
For completeness, we give an outline of the model used to describe the triple system, referring to Ref.~\cite{Santos:2025ass} for details. To model the SB in the MBH spacetime, we exploit the relation $d_{\rm SB} \ll M$ and use Dixon's formalism~\cite{Dixon:1970zza,Dixon:1970zza,Dixon:2015vxa,Costa:2014nta,Harte:2011ku}. Effectively, the SB is described as a point particle with internal structure up to quadrupolar order~\cite{Steinhoff:2009tk,Harte:2020ols}, thus capturing the GWs produced by its internal motion. The spin and quadrupole moment of the SB are tensors in the tangent space to the CoM, whose specific form is in Eq. (S.30) of the Supplemental Material of Ref.~\cite{Santos:2025ass}.

A significant simplification when the SB is held stationary is that now there is a single characteristic frequency, the intrinsic frequency of the SB inner motion:
\begin{equation}
    \Omega_{\rm SB} = \frac{1}{u^t}\sqrt{\frac{M_{\rm SB}}{d_{\rm SB}^3}} \, , \label{eq:frequencies_BL}
\end{equation}
where $u^t$ is the $t$ component of the 4-velocity of the CoM of the SB. Since the later is taken to be stationary, we have $u^t=\left(1-2Mr/ \Sigma\right)^{-1/2}$. To calculate the gravitational perturbation produced by the SB, we solve the Teukolsky equation~\cite{Teukolsky:1972my,Teukolsky:1973ha} in the frequency domain. Requiring the solution to satisfy purely outgoing boundary conditions at future null infinity, as well as purely ingoing on the horizon, it is easy to show that the asymptotic expression for the Weyl scalar is
\begin{equation}
    \Psi_4 \sim \sum_{\substack{\ell, m,q}} {Z}^\infty _{\ell m q} \, _{-2}S^{a } _{ \ell m \omega_{q}} (\theta) e^{i m \phi} \frac{e^{-i\omega_{q} u}}{r} \, , \label{eq:psi4_inf}
\end{equation}
where $\{u,r,\theta,\phi \}$ are (retarded) Boyer-Lindquist coordinates in the MBH spacetime,  and $ _{-2}S^{} _{ \ell m \omega}$  are the $s=-2$ spin-weighted spheroidal harmonics~\cite{Press:1973zz,1986JMP....27.1238L} with spheroidicity $a\omega$, angular mode number $\ell\geq2$, and azimuthal mode number $m$ in the range $-\ell \leq m \leq \ell$. The frequencies excited by the system are 
\begin{equation}
    \pm \omega \equiv \omega_{q} = q \, \rm \Omega_{\rm SB} \, , \label{eq:frequencies}
\end{equation}
where $q\in\{\pm2\}$ is the quadrupole mode number. The range of $q$ reflects the quadrupole approximation taken for the source. 

In comparison with the full model developed in Ref.~\cite{Santos:2025ass}, the frequencies are now only labeled by the quadrupole number, and the GW signal is monochromatic. This is precisely because the center of mass is held static; otherwise, $\omega$ would also have contributions due to the orbital timescale and to geodesic precession~\cite{vandeMeent:2019cam,Rindler_Perlick_1990}. The amplitudes ${Z}^\infty _{\ell m q}$ take the form
\begin{equation}
    Z_{\ell m q}^\infty = \sum_ 
    {i= 0} ^{4}\ \sum_ 
    {j = 0} ^{4-i} \mathcal{A}^{(i,j)}_{\ell m q} \frac{d^i }{dr^i} R^{H}\left(r_{\rm S} \right) \ \frac{d^j}{d\theta^j}  \bar S \left( \pi/2\right) \, , \label{eq:amplitudes}
\end{equation} 
where $\mathcal{A}$ depends only on the outer and inner orbital parameters of the SB, $R^{H}\equiv R^{H}_{\ell m \omega_{q}}$ is a solution to the homogeneous Teukolsky equation satisfying purely ingoing boundary conditions at the horizon, and $\bar S \equiv {_{-2}\bar{S}_{\ell m \omega_{q}}}$. These eigenfunctions are calculated semi-analytically~\cite{Mano:1996vt,Fujita:2004rb,Leaver:1985ax,Berti:2005gp,vandeMeent:2015lxa}. In a given simulation, for each value of $q$ we must compute the amplitudes $Z_{\ell m q}^\infty$ for many harmonics $(\ell,m)$ until we meet a convergence criterion at $\ell=\ell_{\rm max}\sim \Omega_{\rm SB} \, r_{\rm S} $. This criterion is defined in Eq.~(S.5.B) of the Supplemental material of Ref~\cite{Santos:2025ass}. 

The sum over modes in Eq.~\eqref{eq:psi4_inf} can be made more efficient when the system exhibits a reflection symmetry about the equatorial plane ($\theta=\pi/2$). In that case, the amplitudes satisfy $Z^{H,\infty}_{\ell (-m)(- q)} =(-1)^\ell\bar{Z}^{H,\infty}_{\ell m q} $. We will make use of this symmetry throughout, unless stated otherwise.
 
A simple equation relates the Weyl scalar $\Psi_4$ in Eq.~\eqref{eq:psi4_inf} to the strain at future null infinity in transverse traceless gauge 
\begin{equation}
    \Psi_4 \sim \frac{1}{2} \partial_t^2 h \, , \quad h \equiv (h_+ - i h_\times) \, ,
    \label{eq:strain}
\end{equation}
where $h_{+,\times}$ are the usual linear polarizations of GWs and $h$ is the complex strain. 
%
\subsubsection{The amplification function in the wave optics regime}
%

A key quantity in any lensing analysis is the amplification function
\begin{equation}
        F(\omega; \Theta) = h_{\rm L}(\omega;\Theta)/h_{\rm U}(\omega;\Theta) \, , \label{eq:amp_factor}
\end{equation}
where $h_{\rm L,U}$ are, respectively, the lensed and unlensed complex strains, as defined in Eq.~\eqref{eq:strain}, $\omega$ is the gravitational wave frequency and $\Theta$ are the quantities describing the source-lens-observer geometry. In this setup, we fix the source angles $(\theta_{\rm S}, \phi_{\rm S}) = (\pi/2,0)$, so $\Theta= \{a,r_{\rm S},r_{\rm O},\theta_{\rm O},\phi_{\rm O} \}$. We will focus on the absolute value of the amplification factor, which is just the ratio of sums of the amplitudes in both polarizations for the lensed and unlensed signals. The analysis of polarization mixing in these systems is left for future work.

To construct the amplification function, we need the lensed signal, as well as the unlensed signal from an identical binary in isolation. The lensed signal is obtained using the model described in Sec.~\ref{sec:b_EMRI} and Eq.~\eqref{eq:strain}. For the unlensed signal, we take that of a circular Newtonian binary in isolation with orbital frequency $\Omega_{\rm SB}$. In the quadrupole approximation and using a flat spherical coordinate system $(T, R, \vartheta_{\rm S}, \varphi_{\rm S})$ centered at the binary and with poles along its angular momentum vector, the strain seen by distant observers in TT gauge is 
\begin{equation}
    h_U = A\left(\frac{1+\cos^2 \vartheta_{\rm S}}{2} \cos \xi -i \cos \vartheta_{\rm S} \sin \xi \right)  \, , \label{eq:unlensed}
\end{equation}
where $\xi = \omega U + 2 \varphi_{\rm S}$, for $U=T-R$ is the retarded time and $A=4 M_{\rm SB}^2/R \, d_{\rm SB} $. When calculating $h_{\rm U}$ to use in Eq.~\eqref{eq:amp_factor}, one has to be careful in mapping $(r_{\rm S}, r_{\rm O}, \theta, \phi) \mapsto (R,\vartheta_{\rm S}, \varphi_{\rm S})$.

We are interested in comparing the results obtained with the full model against an approximation obtained in the high frequency limit, using the ray tracing methods developed in Ref.~\cite{Vujeva:2026rqh}. We describe the method to obtain the latter in the following subsection. 
%
\subsection{Lensing model from ray tracing} \label{sec:BGO}
%
The model introduced in the previous section has the advantage of being developed from first principles and holding to all orders in $\omega$. However, it misses much of the intuition provided by the geometric optics approximation. For this reason, and also to serve as a comparison for the WO results, we developed a ray tracing based model (RT) that captures finite frequency effects in Ref.~\cite{Vujeva:2026rqh}. This model is currently implemented for lensing close to perfect alignment in the Schwarzschild geometry. As it turns out, this framework significantly outperformed our expectations (see Sec.~\ref{sec:results}), and ended up effectively bridging the strong/weak and wave/geometric optics regimes.
%
\subsubsection{The Strong-Field Amplification Function}
%
Here we summarize the construction of the strong-field amplification function $F$, defined as in Eq.~\eqref{eq:amp_factor} by the ratio of the lensed to the unlensed strain and refer the reader to Ref.~\cite{Vujeva:2026rqh} for the derivations and for the validation of the construction against the weak-field limit. The idea is to generalize the diffraction integral formalism~\cite{Schneider:1992bmb, Takahashi:2003ix, Tambalo:2022wlm} of weak field lensing to the strong field. That is, we define a strong field amplification function as 
\begin{equation}
    F(\omega;\phi_{\rm O})={\kappa\,\omega}\int_{(b,\varphi)}  b\, \mathrm{d}b \; \mathrm{d}\varphi \;e^{i\omega \; \mathfrak{t}_\mathrm{d}(b,\varphi \, ; \, \phi_{\rm O})} \, ,
    \label{eq:diffrac_rt}
\end{equation}
where $\kappa$ is a normalization constant (see Eq.~\eqref{eq:normalization}), $(b,\varphi)$ are polar coordinates in a plane perpendicular to the optical axis at the observer, and $\mathfrak{t}_\mathrm{d}(b,\varphi \, ; \, \phi_{\rm O})$, the strong field time delay surface; the crucial, non-trivial object. In the usual diffraction integral formalism, the phase has the Fermat potential $T_d$, which encodes the weak field time delay between neighboring rays connecting the observer and the source; the diffraction integral merely computes the interference of said rays. The Fermat potential is usually computed with a weak field lens potential. Here, we generalize the construction to the strong field through $\mathfrak{t}_\mathrm{d}(b,\varphi \, ; \, \phi_{\rm O})$, which again, encodes the relative phase of different rays connecting observer and source. The main difference is that, in the strong field, one must trace exact null geodesics of Schwarzschild spacetime to build this surface. 

Our approach includes no beyond geometric optics effects in the propagation of the radiation from source to observer~\cite{Bruyere:2026xzb,Cusin:2019rmt}. Instead all the finite frequency effects arise from solving the integral in Eq.~\eqref{eq:diffrac_rt} as is, rather than employing the stationary phase approximation. Physically, this amounts to including all orders in the frequency for the interference, but assuming radiation travels attached to null rays. 

The relative arrival time inherits its structure from the strong-field geodesics through the eikonal generating function (otherwise known as the reduced radial action). The details of the presentation and its construction are given in the companion paper~\cite{Vujeva:2026rqh}, but for completeness, we give the result here 
\begin{equation}
    \mathfrak{t}_\mathrm{d}(b,\varphi ; \phi_{\rm O}) = \mathcal{I}(b) - b\,( \phi_{\rm O}-\pi)\cos\varphi -  \mathfrak{t}_\mathrm{d}^\mathrm{ref}\, .
    \label{eq:tsurface}
\end{equation}
In the above, $\mathfrak{t}_\mathrm{d}^\mathrm{ref}$ is a constant which is chosen such that the first ray to arrive does so at $\mathfrak{t}_\mathrm{d}=0$. More importantly, $\mathcal{I}(b)$ is the generating function
\begin{equation}
    \mathcal{I}(b) = T(b) - b\,\Phi(b) = \int p_r\,\mathrm{d}r\,, \label{eq:radial_action}
\end{equation}
where $p_r$ is the radial momentum, $T(b)$ is the accumulated coordinate time, and $\Phi(b)$ the accumulated azimuth. In the notation of~\cite{Vujeva:2026rqh}, $\beta_{\rm S}\equiv \phi_{\rm O}-\pi$.

In practice, $\mathfrak{t}_\mathrm{d}^\mathrm{ref}$ is built in the following way: first, fix the source-lens-observer distances, and consider the case where they are perfectly aligned. Trace rays perpendicularly from the observer, at an impact parameter $b$, until they reach the source the source radius $r_{\rm S}$, tracking the associated accumulated coordinate time $T(b)$ and azimuthal angle $\Phi(b)$. This involves solving an elliptic integral and an ODE. 
Finally, shifting the source infinitesimally away from the azimuthal angle introduces the closed form geometric factor $b\,(\phi_{\rm O} - \pi)\cos\varphi$. In this approximation, Eq.~\eqref{eq:tsurface} precisely yields the relative phase of rays hitting different points in the observer screen. In particular, its stationary points yield the relativistic images usually obtained through the stationary phase approximation.

The near-antipodal placement is deliberate, as it puts the source in the region of maximum amplification. The observer is placed far enough from the lens ($r_{\rm O} > 10^3 M$) to sit well within the asymptotically flat region, so that the results below are representative of the $r_{\rm O}\to\infty$ configuration relevant to extragalactic sources. We work with flat-space distances throughout, but for configurations in which any leg of the source--lens--observer geometry is cosmological, these must be promoted to angular-diameter distances once a cosmology is fixed~\cite{Peebles:1994xt}.

So far, we have left out two ingredients in Eq.~\eqref{eq:diffrac_rt} : the integration region and the normalization. Regarding the first, the RT procedure can only be carried out if the geodesic reaches the source radius and does not plunge into the BH. Thus, the arrival-time surface is well defined only on the connecting annulus $b \in (b_{\rm c}, b_\mathrm{max}]$. The inner edge is the critical impact parameter $b_{\rm c} = 3\sqrt{3}\,M$, and the outer edge is the impact parameter at which the turning point matches the source radius (i.e. $r_{\rm t} = r_{\rm S}$),
\begin{equation}
    b_\mathrm{max}(r_{\rm S})= \sqrt{\frac{r_{\rm S}^3}{r_{\rm S}-2M}}\,.
\end{equation}
Crucially, this finite support sets a minimum $M\omega$ below which the resulting amplification factor cannot be trusted: since the integration region in Eq.~\eqref{eq:diffrac_rt} is finite, very low frequencies lead to uninformative, nearly constant phases.  Expanding $\mathfrak{t}_\mathrm{d}$ to quadratic order
about its minimum at $b_0$ yields $\omega \gtrsim 1/(b_\mathrm{max}-b_0)^2 |\mathcal{I}''(b_0)|$.

As $b\to b_{\rm c}$, geodesics start winding around the MBH before reaching the source~\cite{Gralla:2009md}; we call winding number the amount of winds around the MBH. The arrival-time surface is constructed order by order in the winding number $n$ of the geodesics, and the total amplification factor is the sum of the contributions of each order,
\begin{equation}
    F_{\rm tot}(\omega) = \sum_{n} F^{(n)}=(\omega)\,.
    \label{eq:Ftot}
\end{equation}
where $n$ is the winding number of the geodesic. Because the higher-winding images are exponentially de-magnified \cite{Bozza:2001xd, Bozza:2002zj}, in this work we only compare the direct ($n=0$) term when comparing against the full wave scattering results and drop the $n$ superscript on $F$, unless stated otherwise.

Finally, $\kappa$ is the normalization constant of the integral. We obtain it from $\mu$, the geometric-optics magnification of an individual image. The latter is distinct from the amplification function $F(\omega)$, which is a frequency-dependent complex quantity encoding the coherent sum over the whole screen. The two are connected only in the high-frequency limit, where the stationary phase approximation reduces $F$ to a sum over images weighted by $\sqrt{|\mu|}$. This correspondence, in turn, fixes
\begin{equation}
  \kappa \;=\; \frac{1}{2\pi i}\sqrt{\frac{\mu^\mathrm{ref}\,|\beta_S|\,|\mathcal{I}''(b^\mathrm{ref})|}{b^\mathrm{ref}}}\;, \label{eq:normalization}
\end{equation}
where $\mu^\mathrm{ref}$ is the magnification of the image at the minimum of the time delay surface (i.e. the image that arrives first), and
$b^\mathrm{ref}$ its corresponding impact parameter. We refer the reader to Ref.~\cite{Vujeva:2026rqh} for the matching.

\subsubsection{Quadrupolar Emission Effects}
\label{sec:spin2}

The phase carried by a geodesic is set by the direction in which it is emitted, measured in the source's local rest frame. For the static Schwarzschild source considered here, that frame is the static orthonormal tetrad at the emission point. Projecting the photon four-momentum $k^\mu$ at the intersection onto this tetrad, we define the emission angle as 
%
\begin{equation}
 \varphi_{\rm S}(b,\varphi)=\mathrm \arctan2\! \ \big(\cos\psi,\,\sin\psi\cos\varphi\big),
    \label{eq:emangle}
\end{equation}
where $\psi$ is the local emission angle measured from the outward radial direction in the source's frame, with ($\psi=\pi-\arcsin[b\sqrt{1-2M/r_{\rm S}}/r_{\rm S}]$), and $\varphi$ is the observer-screen azimuthal angle. We emphasize that $\varphi_{\rm S}$ is the angle in the source's own frame and not the angle projected from the observer screen. Using the latter would misassign the phase. 

For a quadrupolar source in the equatorial plane, the emission carries the phase $e^{2i\varphi_{\rm S}}$, with the factor of two being the imprint of the quadrupole (identical to that in the toy model of Appendix~\ref{ap:toymodelsource}), and the diffraction integral per winding number becomes
\begin{equation}
    {F}(\omega) = {\kappa\omega}\int_{(b,\varphi)}b\,  \mathrm{d}b \; \mathrm{d}\varphi\;
    e^{\,2i\varphi_\mathrm{S}(b,\varphi)}\;e^{i\omega\,\mathfrak{t}_\mathrm{d}(b,\varphi)}\,.
    \label{eq:diffrac_spin2}
\end{equation}

Throughout this work, we refer to $F_0$ as the scalar amplification function (Eq.~\ref{eq:diffrac_rt}), familiar from the weak-field literature, built for a single point-source position, and not taking the quadrupolar emission into account. 
We denote the amplification function containing the quadrupolar structure as $F_2$. Note the difference in notation to the companion paper~\cite{Vujeva:2026rqh} where $F_n$ denote different winding numbers.

%
\section{Results}\label{sec:results}
%
We will mostly present the results for a fiducial set of system parameters, given in Tab.~\ref{tab:parameters_fiducial}. The spin of MBH $a$ will be 0 in Sec.~\ref{sec:results_Schwarzschild} and $0.9 M$ in Sec.~\ref{sec:results_Kerr}. The CoM of the SB is stationary on the $x$ axis, at $\theta_{\rm S}=\pi/2$, $\phi_{\rm S}=0$. 
%
\subsection{Non rotating massive black hole} \label{sec:results_Schwarzschild}
%
We start with the case where the MBH is not rotating, focusing on the two SB spin orientations, labeled by SB I and SB II, in Fig.~\ref{fig:binary_geometry}. They correspond to the SB spin pointing at the MBH or perpendicular to it. 

\begin{figure}
    \centering
    \includegraphics[width= \linewidth]{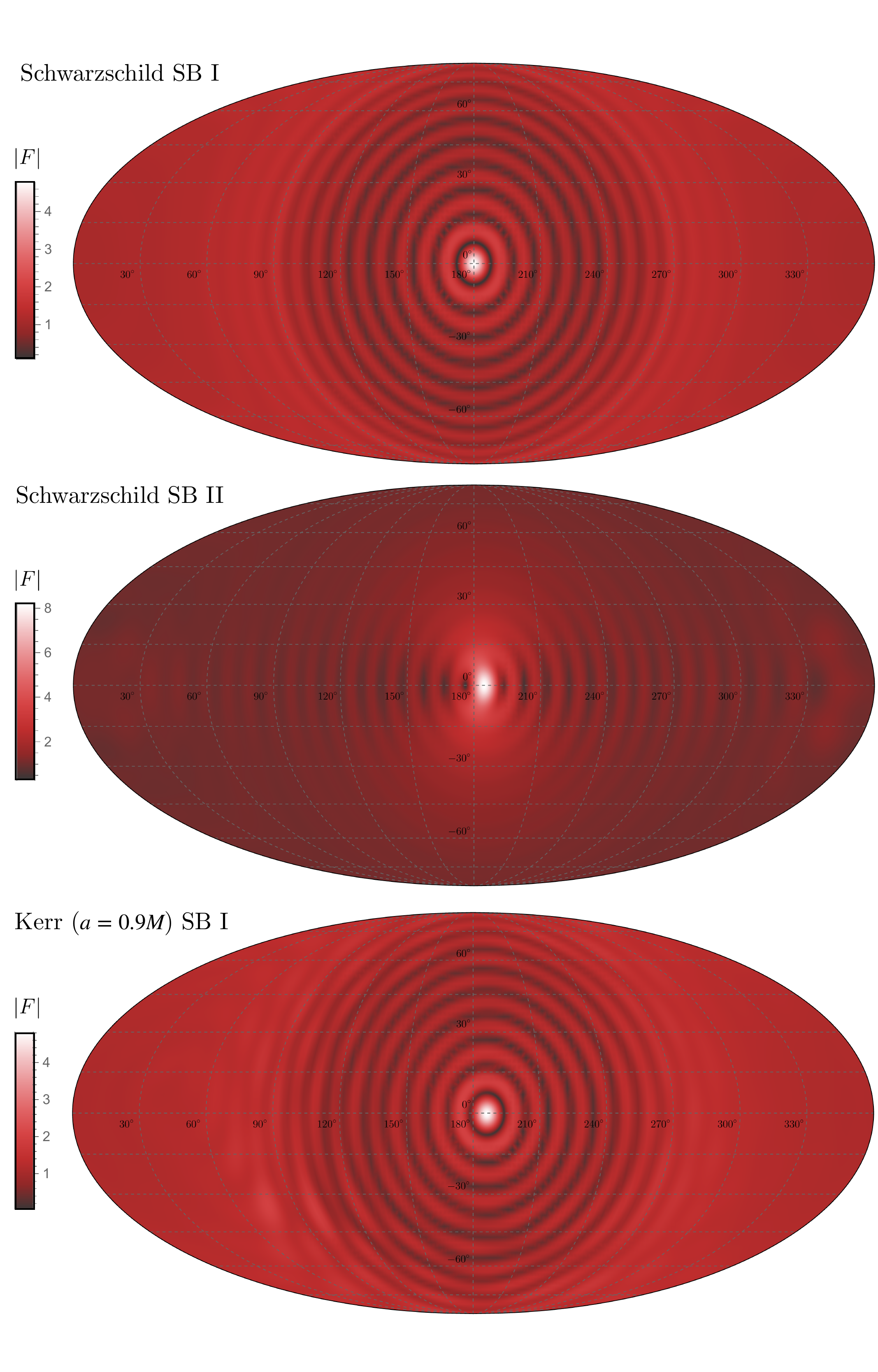}
    \caption{Mollweide projections of the absolute value of the amplification factor $|F|$, computed using the fully relativistic wave optics (WO) model (see Sec.~\ref{sec:b_EMRI}), as a function of the observer coordinates in the sphere at future null infinity ($\theta_{\rm O}$ and $\phi_{\rm O}$). The source is a stationary binary, with the parameters of Tab.~\ref{tab:parameters_fiducial}. \textbf{Top:} The lens is a Schwarzschild MBH and the spin of the SB is pointing at the lens (SB I in Fig.~\ref{fig:binary_geometry}). \textbf{Middle:} The lens is a Schwarzschild MBH and the spin of the SB is pointing in the $\hat{z}$ direction (SB II in Fig.~\ref{fig:binary_geometry}). \textbf{Bottom:} The lens is a Kerr BH with $a=0.9 M$ pointing in the $\hat z$ direction, and the spin of the SB is pointing at the lens (SB I in Fig.~\ref{fig:binary_geometry}). We find the fringes characteristic of WO. The amplification peaks when the observer is at (near) the antipodal point in the configuration of the top (middle and bottom) panel(s). } 
    \label{fig:Mollweides}
\end{figure}

We can get a general picture of lensing in this regime by looking at the amplification factors for an observer at future null infinity, $\mathscr I^+$ (in the limit $r_{\rm O} \to \infty$). Fixing the fiducial parameters and the MBH spin, we get $F=F(\theta_{\rm O}$, $\phi_{\rm O}$), so we plot it on the celestial sphere using a Mollweide projection~\cite{weisstein}. The results for the fiducial parameters (Tab.~\ref{tab:parameters_fiducial}) are in the top two panels of Fig.~\ref{fig:Mollweides}. 

We find fringes in all cases, resulting from consecutive regions of constructive and destructive interference. For SB I, maximum amplification occurs at the antipodal point ($\theta_{\rm O}=\pi/2$, $\phi_{\rm O}=\pi$). Contrarily, for SB II, maximum amplification occurs close to but not at the antipodal point. This shift can be mistaken for numerical error; however it is the result of the (non isotropic) time-varying quadrupolar emission profile of the SB. In other words, the source-lens-geometry is not axisymmetric in SB II, which is why we don't get an axisymmetric diffraction pattern. This is discussed in more detail using a toy model in Appendix~\ref{ap:toymodelsource}\footnote{A similar result has been experimentally verified for electromagnetic radiation~\cite{doi:10.1126/science.1233739}.}. SB II also has a higher maximum amplification, simply because in this setup the unlensed signal is weaker near the antipodal point (c.f. Eq.~\eqref{eq:unlensed}). 

The exact fringe spacing and amplifications depend on more details of the setup, namely the frequency  $\omega$ and the source position $r_{\rm S}$. Next, we quantify these dependencies, looking at SB I and SB II individually. We focus on the results near perfect alignment (in the region of maximum amplification). The reason for this is twofold: \textit{(i)} this is the region with highest observational prospects \textit{(ii)} this region is near the caustic of the lensing setup~\cite{Ezquiaga:2025gkd}, where there are universality properties we want to test in the WO regime.
%
%
\subsubsection{SB I : axisymmetric source-lens setup} \label{sec:SBI}
\begin{figure}[h]
    \centering 
    \includegraphics[width=0.9\linewidth]{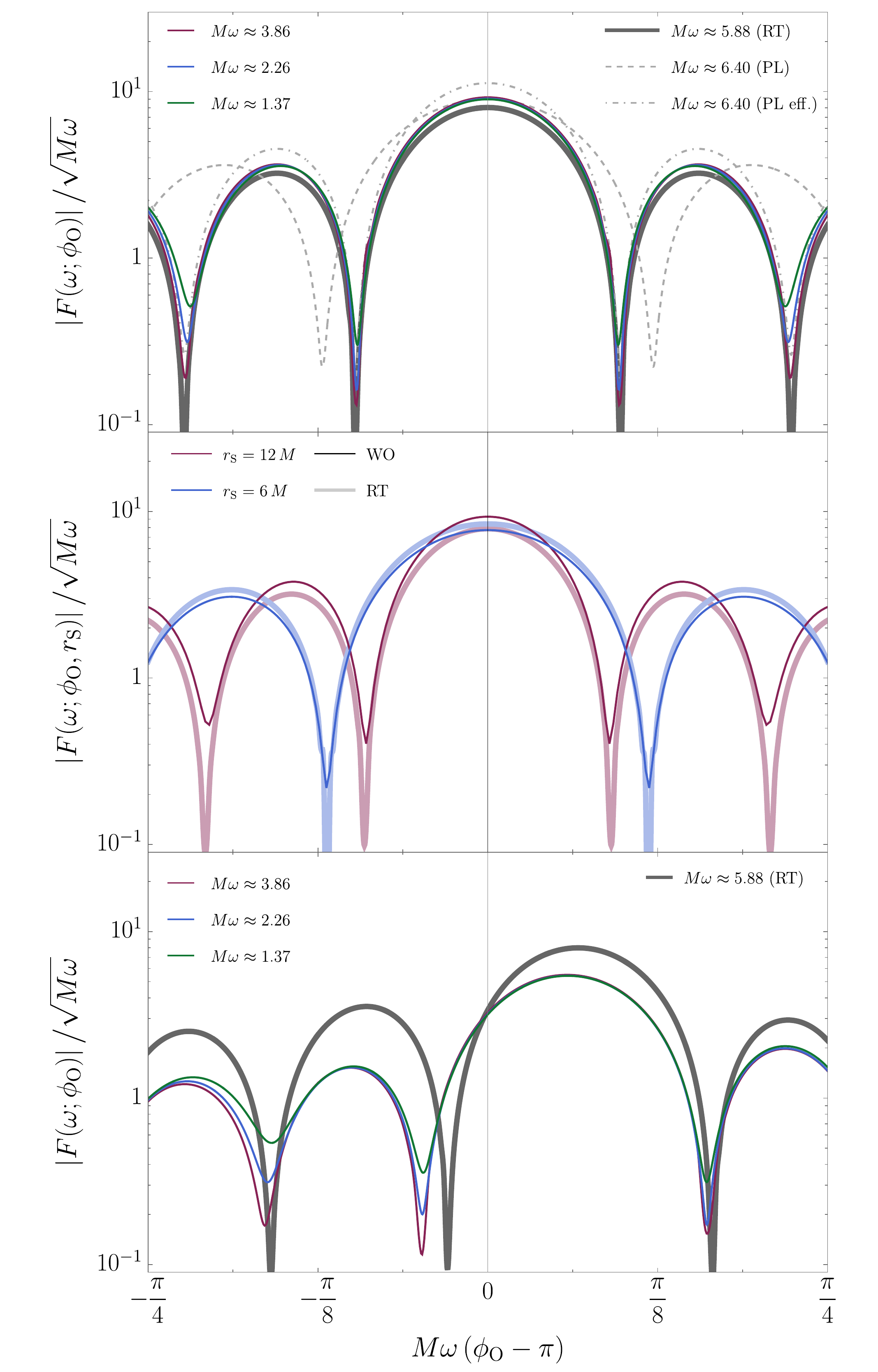}
    \caption{Amplification function for an equatorial observer ($({\theta_{\rm O}},{\phi_{\rm O}})=(\pi/2,\phi_{\rm O} )$), in different configurations of the binary near a Schwarzschild MBH (see  Fig.~\ref{fig:binary_geometry}). Thin solid lines are to full wave optics (WO) results; thick lines are obtained with the ray tracing model (RT). Axes are rescaled by appropriate powers of $\omega$. 
    \textbf{Top:} SB I at $r_{\rm S} = 10 M$, for different source frequencies $\omega$. Outstanding agreement between WO curves: peak amplification and first fringe spacing both agree to $\sim$1\%. Comparing to RT, we get $\sim$10 \% agreement, while a point lens (PL) always yields $>20$ \% deviation in either the peak or fringe spacing.
     \textbf{Middle:} SB I with $M \omega \approx 2.12 $, for different source radii. Similar agreement to the above. 
    \textbf{Bottom:} SB II at $r_{\rm S} = 10 M$, for different source frequencies. Maximum amplification does not coincide with perfect alignment due to the breaking of axisymmetry; RT results are able to predict this. RT results predict the peak shift at $\sim 15\%$ level, but the peak amplitude and fringes deviate by $\sim 50\%$.
    All WO results have a universal dependence on the frequency (illustrated by the rescaling of the axes). RT model can describe these results, while a PL model proves inadequate for these sources.
    }
    \label{fig:F_schwarzschild}
\end{figure}

%
We start by focusing on the simpler axisymmetric setup SB I. In the top panel of Fig.~\ref{fig:F_schwarzschild} we plot the amplification factor on the equatorial plane near the antipodal point, comparing different frequencies $\omega$. All curves correspond to placing the binary at $r_{\rm S} = 10 M$, so the lensing geometry is always the same. 

In a weak lensing, thin lens, WO calculation~\cite{Ezquiaga:2025gkd}, a source near a caustic of a spherically symmetric lens produces an amplification factor $F(\omega)\sim \sqrt{M \omega }\times f(M \omega \, (\phi_{\rm O}-\pi))$. For this reason, in all panels of Fig.~\ref{fig:F_schwarzschild}, we rescale the axes accordingly. We find that, even if our setup breaks the weak lensing and thin lens approximations, this rescaling is enough to line the curves up very well. We conclude that, even for $M\omega \gtrsim 1$, the amplification factor already exhibits a universal behavior upon varying $\omega$. This result a direct consequence of limiting the analysis to near perfect source-lens-observer alignment, as it means the source is always close to the caustic~\cite{Bozza:2002zj}. Indeed, we confirmed that if one looks at $M \omega< 1$ or $|\phi_0 - \pi|\gtrsim 1$, the caustic structure is lost, and with it the universality is broken. 

This universality motivated the question: \emph{can we develop an alternative model, based on the simplifying assumptions of the geometric optics limit, that can reproduce this universal behavior at finite frequency?} This question led to the development of the RT model summarized in Sec.~\ref{sec:BGO} and described in detail in the companion paper~\cite{Vujeva:2026rqh}. The RT model is more accurate for higher frequencies, so we compare the WO results with RT evaluated at a higher frequency, always applying the rescaling mentioned in the previous paragraph. The higher frequency RT (solid thick lines in Fig.~\ref{fig:F_schwarzschild}) agrees with the WO results to $\sim10\%$ accuracy, answering affirmatively to the question posed above.

This is a nontrivial result: the thin solid lines are obtained from first principles, with no approximations to the wave propagation besides using linear BH perturbation theory. The gray solid curve results from evaluating the diffraction integral by tracing null geodesics to construct the strong field time delay surface. The fact that they agree by performing a simple frequency rescaling is remarkable. It shows that the considerations of geometric optics, that waves travel along null geodesics, are still physically meaningful and predictive, even for $M \omega\sim 1$. 

We also include in Fig.~\ref{fig:F_schwarzschild} a comparison with the well known Newtonian and analytical point lens model (PL). This model has recently been extrapolated to strong lensing setups, similar to the ones studied here~\cite{Liu:2023ikc,Ubach:2025oxr, Leong:2024nnx,Martens:2026mku}. Our results show that the PL model can indeed reproduce the fringe spacings \emph{or} the maximum amplification near alignment, but not both. Moreover, using this model leads to significantly biased lens parameters. We comment on this in detail in Sec.~\ref{sec:point_lens}.

So far, we have kept the source at $r_{\rm S} = 10 M$. The effect of the source-lens distance on the amplification function is not as straightforward as that of the frequency. Investigating eq.~\eqref{eq:diffrac_rt}, while $\omega$ only enters the diffraction integral through the the phase, $r_{\rm S}$ changes the time delay surface. Thus, no simple rescaling can produce a universal curve in the strong field. Still, we can look at different source positions and compare results from WO and RT. We do this in the middle panel of Fig.~\ref{fig:F_schwarzschild}, keeping all parameters fiducial (see Tab.~\ref{tab:parameters_fiducial}) except $r_{\rm S}$. 
Qualitatively, for increasing $r_{\rm S}$ the diffraction pattern compresses, in line with the weak field literature~\cite{Schneider:1992bmb, Ezquiaga:2025gkd}. As before, we compare with RT results for sources at the same $r_{\rm S}$ but at a higher frequency ($M\omega \approx 4,28$ for $r_{\rm S}/M= 12, 6$, respectively). After rescaling, the WO and RT curves agree remarkably well, even for $r_{\rm S}=6M$. Again, it is the universality of lensing near the caustic that leads to this outstanding agreement. 
%
\subsubsection{SB II : non-axisymmetric source-lens setup} \label{sec:SBII}
%
We turn our attention to the setup labeled SB II in Fig.~\ref{fig:binary_geometry}, where the spin of the SB is perpendicular to the radial axis connecting it to the MBH. In the bottom panel of Fig.~\ref{fig:F_schwarzschild} we perform the same analysis as for SB I in the top panel. A noteworthy difference is that, in this setup, axisymmetry breaking implies that evaluating $F$ along a meridian, instead of along the equator, will lead to different results (see Fig.~\ref{fig:Mollweides}). Continuing to look at the equator is, at this level, a choice. 

As mentioned already, the maximum amplification does not occur at the antipodal point due to the breaking of axisymmetry. Still, just as in the case of SB I, we find that different values of $\omega$ lead to (after appropriate rescaling)  universal amplification factors. As before, this universality can be attributed to the proximity to a point caustic in the lensing problem.

The agreement after rescaling is not as good as for SB I, though the curves are clearly converging to their leading high frequency component. This motivates, once more, trying to recover this limiting curve using a RT approach. However, this is more intricate for the SB II setup because the interfering phases in the diffraction integral must now account for an extra phase difference at emission, adding to the one caused by the time delay (see Eq.~\eqref{eq:diffrac_spin2}). This is best illustrated in the toy model presented in Appendix~\ref{ap:toymodelsource}. This feature has been implemented in the RT model, and the result is the gray curve in Fig.~\ref{fig:F_schwarzschild}. 

The RT model agrees moderately well with the WO curves. In particular, it predicts the angular profile of $F$, including the axisymmetry breaking. Still, the agreement is not at the level of what we found for SBI (top panel). This is likely explained by the simple implementation of the quadrupole nature of the source. For example, it currently only includes the phase different between different emission directions, but not the emission power anisotropy nor the existence of two independent helicity modes. Improving the source emission model is a natural extension of the RT model. With proper source modeling, there is no reason to believe that the level of agreement observed in the top panel of Fig.~\ref{fig:F_schwarzschild} cannot be achieved.
%
\subsection{Rotating massive black hole} \label{sec:results_Kerr}
%
Finally, we investigate the case where the lens is a spinning Kerr MBH with spin $a=0.9 M$. The phenomenology of lensing by a Kerr BH is far richer than for non-spinning lenses, and has been extensively explored in the literature~\cite{Bozza:2010xqn, Bozza:2008mi, Feleppa:2026emm, Sereno:2007gd}.

In the bottom panel of Fig.~\ref{fig:Mollweides}, we show the amplification factor on the celestial sphere for a fiducial (see Tab.~\ref{tab:parameters_fiducial}) system in a Kerr spacetime with $a=0.9 M$. First thing to note is that, even in the more symmetric setup of SB I (see Fig.~\ref{fig:binary_geometry}), we no longer get maximum amplification for perfectly aligned source-lens-observer geometry. This is due to the frame dragging effect in a rotating BH spacetime. Specifically, in the geometric optics limit, we expect maximum amplification (formally divergent) to occur when the source crosses the \emph{caustic}: defined in the source plane as the set of points where the mapping from this plane to the observer becomes singular. The caustic maps to the critical curve in the lens plane. Given the spherical symmetry of Schwarzchild spacetime, critical curves for all observing angles are circular. All points along the critical curve map back to one single point on a source plane, generating a \emph{point caustic}. This is a generic property of 1 parameter spherically symmetric lens mappings, and a Newtonian analog can be seen in the point mass lens approximation~\cite{Ezquiaga:2025gkd}. However, the introduction of the spin parameter in Kerr breaks this symmetry, particularly for equatorial (or near-equatorial) observers. The deformation of the critical curve due to the frame dragging effects causes the point caustic to be deformed into an astroid~\cite{Bozza:2008mi}. This astroidal shape is common in nearly all two parameter families of lenses, and are comprised of two types of caustics: folds (edges) and cusps (vertices)~\cite{Schneider:1992bmb, Ezquiaga:2025gkd}. A more detailed description of the caustics of Kerr spacetimes can be found in Refs.~\cite{Bozza:2002af, Sereno:2007gd, Bozza:2008mi, Bozza:2010xqn, Feleppa:2026emm}. 

\begin{figure}
    \centering
 	\includegraphics[width=\linewidth]{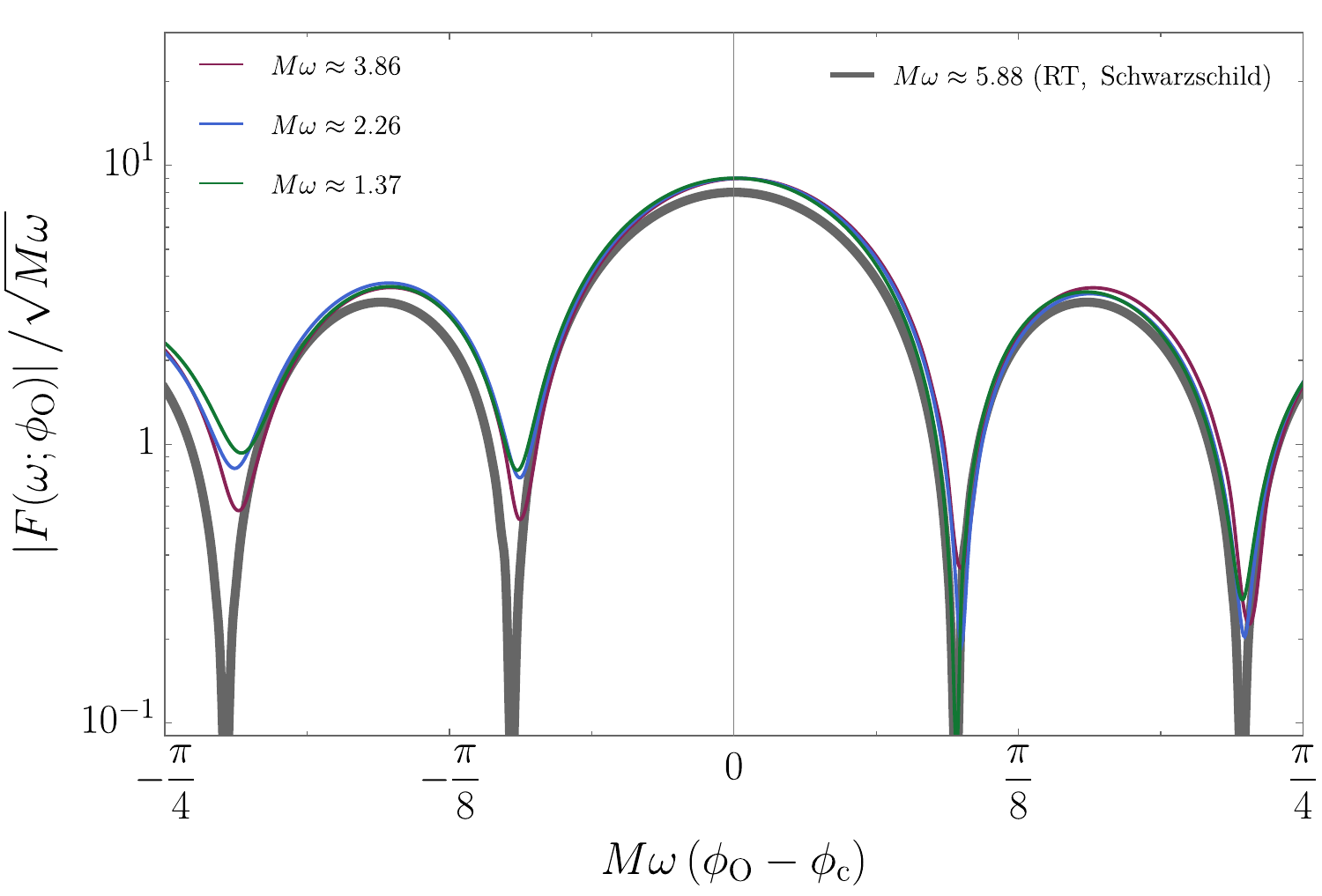}
    \caption{\justifying Same as the top panel of Fig.~\ref{fig:F_schwarzschild} but for a massive Kerr BH with $a=0.9 M$ as the lens. After rescaling, the amplification factor has a universal dependence on the observer angle. The maximum amplification does not coincide with perfect alignment, but instead with location of the caustic tube at the source radius $\phi_{\rm O} = \phi_{\rm c}$ (see Eq.~\eqref{eq:caustic}). Remarkably, the RT results in Schwarzschild (the same as in Fig.~\ref{fig:F_schwarzschild}), are in very good agreement with the rescaled WO results in Kerr. 
    }
    \label{fig:F_Kerr_SBI}
\end{figure}

In Fig.~\ref{fig:F_Kerr_SBI}, we compare the amplification factors for sources near a Kerr MBH with $a=0.9 M$. The sources are in the SB I configuration (see Fig.~\ref{fig:binary_geometry}), we vary their frequency, and keep the other parameters the same as in Tab.~\ref{tab:parameters_fiducial}. We perform an identical rescaling of the curves with the frequency as in the Schwarzschild BH case, but instead show azimuthal shifts with respect to the caustic angle $\phi_{\rm c}$ (the details of which can be found in Appendix~\ref{ap:kerrcaust}). 

We compare these with the RT results in Schwarzschild (the same as in the top panel of Fig.~\ref{fig:F_schwarzschild}). We find surprisingly good agreement between the full WO and RT results. Note that the RT model can be straightforwardly generalized to Kerr, with the problem that mapping out the time delay surface becomes significantly more expensive. Still, these results show that RT can still inform the WO lensing problem in Kerr, which is, at least in the equatorial plane, similar to that of Schwarzschild up to an azimuthal shift of the caustic. This statement holds for sources at $r_{\rm S}=10M$ and spins $0<a/M<0.95$.

The agreement between the Kerr WO and Schwarzchild RT results stems from the fact that we simply do not have the frequency resolution to resolve the small-scale structure of the caustic~\cite{Bozza:2008mi}. Thus, after taking the azimuthal shift of the caustics into account, at these frequencies, the true astroidal caustic structure of Kerr is left unresolved and smears into a point-like caustic, reminiscent of the Schwarzchild case. This implies that it would be incredibly challenging to try and constrain the spin of a Kerr BH using lensed GWs in this frequency range.

The results we have presented here have three clear messages about strong field lensing of a stationary SB source near a MBH: \emph{(i) Even at $M \omega \sim 1$, the dependence of the amplification factor on the frequency follows a universality relation, predicted by catastrophe theory; (ii) A RT model (see Sec.~\ref{sec:BGO}) can reproduce the amplification in this region to a very good degree; (iii) In the strong field, the quadrupole emission of the SB significantly impacts the angular shape of the diffraction pattern.}
%
 \subsection{Extrapolating point lenses} \label{sec:point_lens}

In this section we derive the correction terms to the PL Einstein radius in order to approximate the diffraction patterns found in the full scattering problem and the RT approach in the strong field.  A key reason why the PL produces different diffractive behavior from the strong field approaches is the mismatch between the predicted relative arrival times of the images, accompanied by an underestimation of their separation near alignment. Following Ref.~\cite{Keeton:2005jd}, we can write the asymptotic deflection angle $\alpha(b)$ through cubic order in $M/b$ as

\begin{equation}
\alpha(b)=\frac{4M}{b}+\frac{15\pi}{4}\frac{M^{2}}{b^{2}}
          +\frac{128}{3}\frac{M^{3}}{b^{3}}
          {+\mathcal{O}[(M/b)^4]}.
\label{eq:alpha}
\end{equation}

Using this corrected deflection angle, we can try and relate this back to an effective PL mass or Einstein radius that could reproduce such deflections. Under the thin lens and small-angle assumptions, the ring radius for a source exactly behind the lens satisfies
\begin{equation}
b_\mathrm{E}=\mathcal{D}\,\alpha(b_\mathrm{E}),\qquad
\mathcal{D}\equiv\frac{D_\mathrm{L}D_\mathrm{LS}}{D_\mathrm{S}}
=\frac{r_\mathrm{O}r_\mathrm{S}}{r_\mathrm{O}+r_\mathrm{S}}.
\label{eq:ring}
\end{equation}

Keeping only the leading order term $\alpha=4M/b$ in \eqref{eq:ring} gives us the standard Einstein radius, $b_\mathrm{E,0PN}^{2}=4M\mathcal{D}$.  Defining the dimensionless mass rescaling using $M_{\rm eff}\equiv({b_\mathrm{E}}/{b_\mathrm{E,0PN}})^{2}$ and substituting \eqref{eq:alpha} into \eqref{eq:ring}, we find the implicit expression for the mass rescaling, retaining two corrections beyond the leading deflection term (2PN), to be
\begin{equation}
M_{\rm eff}
=1+\frac{15\pi}{16}\frac{M}{b_\mathrm{E}}
 +\frac{32}{3}\Big(\frac{M}{b_\mathrm{E}}\Big)^{2}
 +\mathcal{O}[(M/b_\mathrm{E})^3].
\label{eq:meff}
\end{equation}

\begin{figure}
    \centering
    \includegraphics[width=\columnwidth]{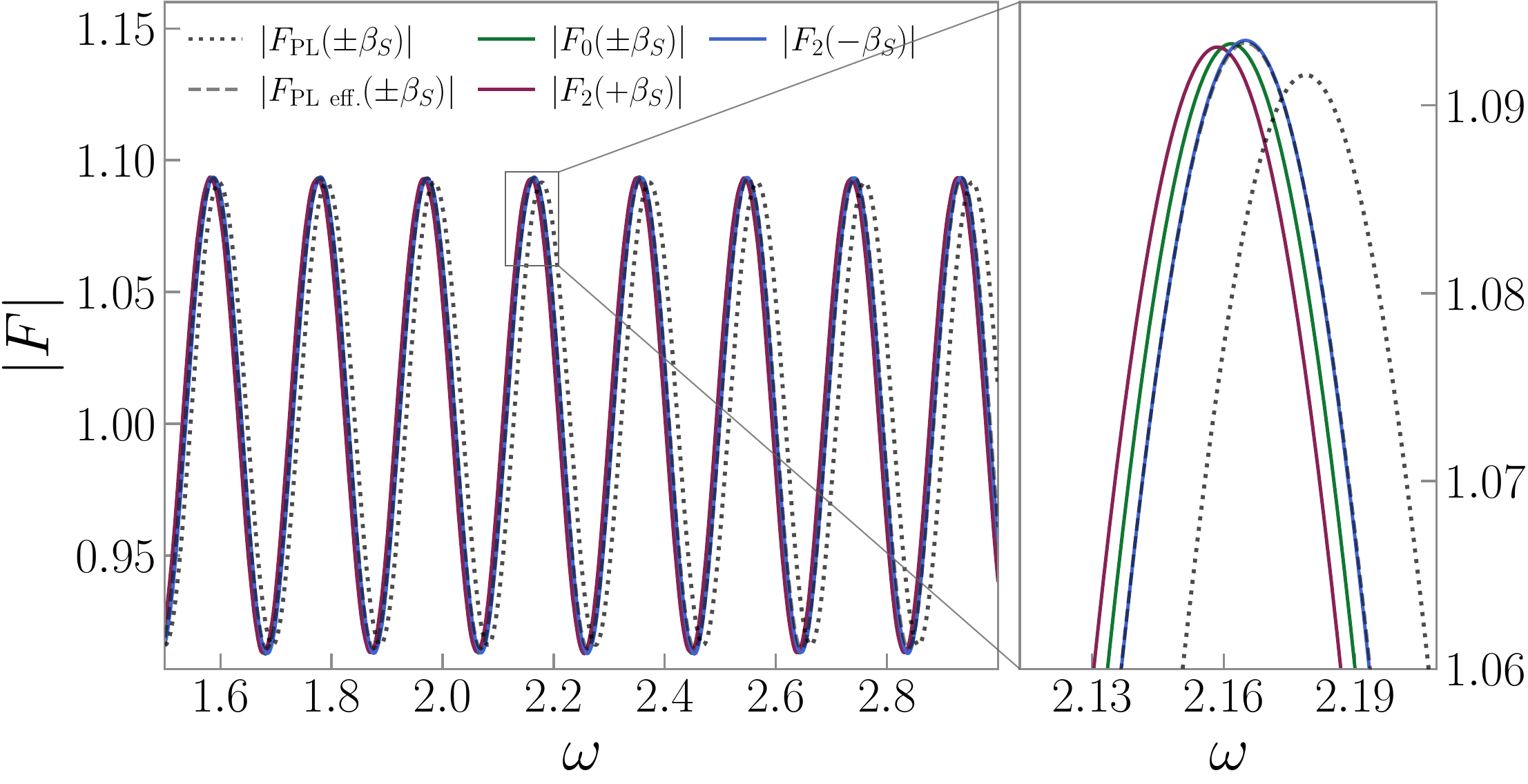}
    \caption{Weak field comparison between the ray traced model, and the standard and corrected PL for a source at $r_O =r_S = 10^4 M$, $\beta_S = \phi_\mathrm{O} - \pi = \pi/36$. Note that even at this distance, there is considerable dephasing between the PL and the effective PL (see Fig.~\ref{fig:meff} for details).
    }
    \label{fig:pl_corrected_comp}
\end{figure}

We can use this form to show why the correction decreases in the weak field.
For $r_\mathrm{O}\gg r_\mathrm{S}\gg M$, $b_\mathrm{E,0PN}\to2\sqrt{Mr_\mathrm{S}}$, so to leading order
\begin{equation}
M_{\rm eff}-1\;\simeq\;\frac{15\pi}{32}\sqrt{\frac{M}{r_\mathrm{S}}}
\;\approx\;1.473\sqrt{\frac{M}{r_\mathrm{S}}}.
\label{eq:sqrtlaw}
\end{equation}
Interestingly, the correction only falls off as $\sqrt{M/r_\mathrm{S}}$. In this distant-observer limit, it is $4.7\%$ at $r_\mathrm{S}=10^{3}M$ and reaches $1\%$ only near $r_\mathrm{S}\simeq2.2\times10^{4}M$. For the finite observer distance $r_\mathrm{O}=10^3M$ used here, the correction at $r_\mathrm{S}=10^3M$ is instead approximately $7\%$. The crucial implications are that, even for conservative estimates of the distance at which the migration traps exist ($10^2 - 10^3 M$), and where these systems form, there are non-trivial corrections to the PL that must be taken into account. This can be seen explicitly in Fig.~\ref{fig:pl_corrected_comp}, where even for a source at $r_\mathrm{S} = 10^4 M$, there is still some dephasing in the amplification function, which although mild at low frequencies, accumulates at high frequencies.
\begin{figure}
    \centering
 	\includegraphics[width=\linewidth]{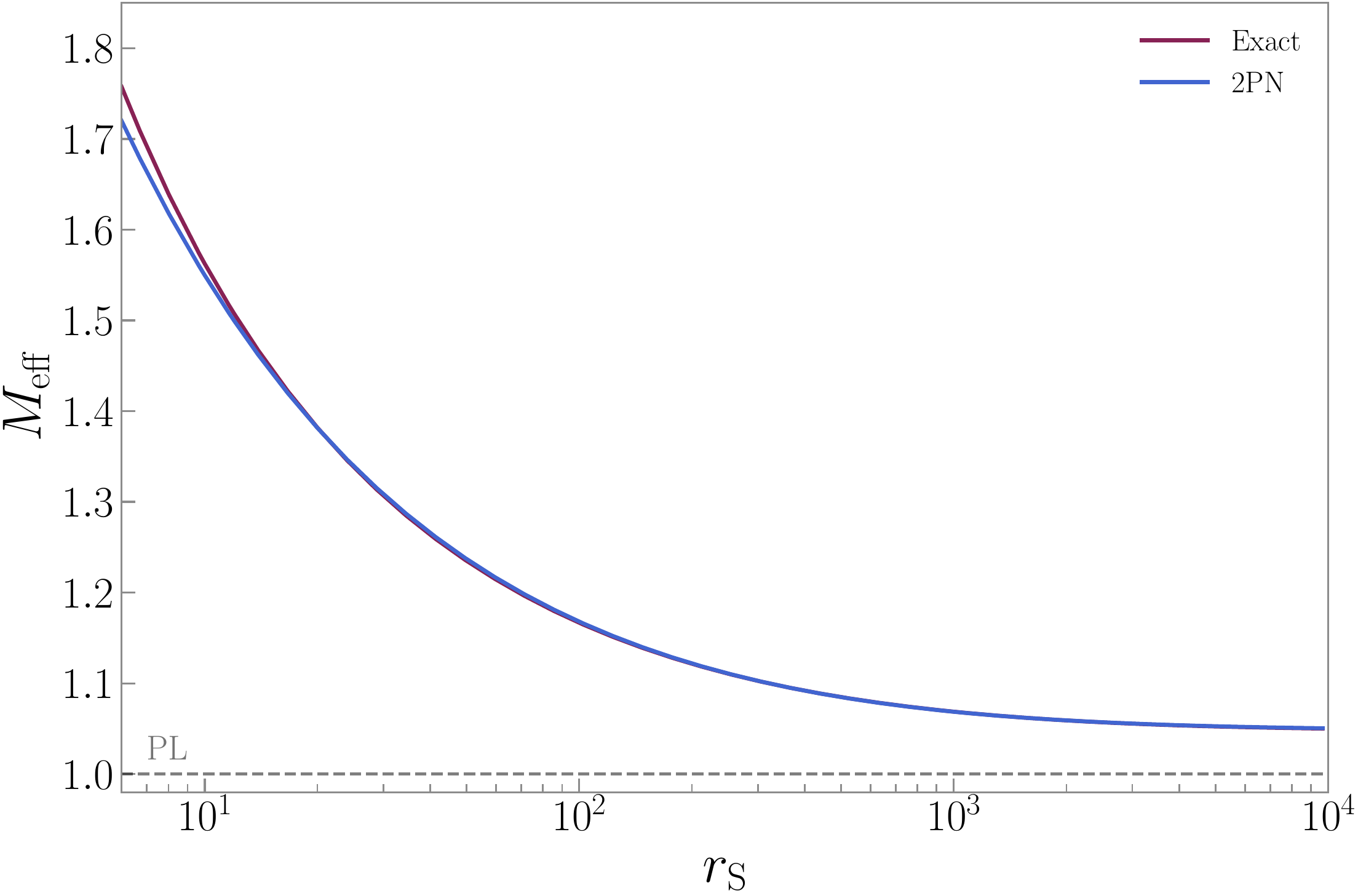}
    \caption{ Effective mass as a function of the distance between the source and the lens, for an observer at $r_\mathrm{O} = 10^3 M$. An $m_\mathrm{eff}$ constructed using fitted image locations and arrival times using the full ray tracing setup is shown for comparison. 
    }
    \label{fig:meff}
\end{figure}

To illustrate the effective mass' dependence on the source distance required to reproduce the fringe spacing in the top panel of Fig.~\ref{fig:F_schwarzschild}, Fig.~\ref{fig:meff} shows the dependence of the effective mass on the source distance for $r_\mathrm{O}=10^3M$. At fixed observer distance, the PN2 rescaling approaches $M_{\rm eff}\simeq1.048$ as $r_\mathrm{S}\to\infty$, remaining within $\sim5\%$ of the PL. Placing the observer at infinity allows agreement to improve to $\sim1\%$ near $r_\mathrm{S}=2.2\times10^4M$. While the corrections decrease at large source distances, we see that for the majority of astrophysically plausible radii for the migration traps that could produce the binaries considered in this work, there is a non-negligible correction to the PL that must be accounted for.
Of equal importance is the note of the limitations of such an effective prescription. Firstly, matching the Einstein radius does not by itself ensure agreement in the relative arrival times. With the other parameters fixed, this rescaling cannot in general match both the magnifications and the arrival time differences of the full scattering or RT approaches in the strong field. In this work, the rescaling was used to show the agreement of the diffractive spacings of the amplification function (c.f. Fig.~\ref{fig:F_schwarzschild}). However, it cannot fully recover the amplitude of the strong field approaches (we find variation of 20-30\% at peak amplification) with only the PL mass rescaled. In other words, a choice must be made between sacrificing accurate amplitudes or arrival times. The corrected PL does not reproduce both in these comparisons.

Additionally, other physical effects expected to be present in such systems (such as the motion and emission profile of the source, or the impact of the lensing on the polarization content of the source) become challenging to implement in the Newtonian framework, which will miss crucial information that is present in the problem~\cite{Santos:2025ass}. Therefore, building upon frameworks that can capture the intricacies of these strong field triple systems will be of utmost importance if we wish to both identify and detect such systems in the future. Additionally, the only regime examined here in which the PL (be it the standard or effective version) can reproduce some of the results of the full WO calculation is in the case of SBI for an equatorial source, which is a significantly less astrophysically plausible source-lens configuration than SBII~\cite{Li:2022jbj, Dittmann:2023sha}. As such, all of the additional effects of having a spin-2 source in this work, as well as further effects such as motion, polarization mixing, precession, etc. will be entirely missed by the PL model. Even in the best case scenario of an equatorial SBI configuration, using the standard PL can bias the mass estimate due to the effects shown in Sec.~\ref{sec:point_lens}. In the worst case scenario, the PL can miss characteristic features in the full WO lensed waveform, which could lead to mis-identifying the event as not coming from a triple system.

In summary, whilst the rescaled PL can reproduce some of the basic features of the strong field methods, it is merely a patchwork solution that neglects much of the interesting physics present in this problem. Notably, the effects of the strong field geometry, the effects of the spin-2 nature of the source, and future improvements of the RT model that will include the effects of polarization mixing, and motion.

%
\section{Conclusions}\label{sec:conclusions}
%

In this work we have studied the wave-optics lensing of the radiation emitted by a stellar binary in the vicinity of an MBH. By comparing a fully relativistic black-hole perturbation theory calculation with the ray-traced beyond-geometric-optics construction described in a companion paper ~\cite{Vujeva:2026rqh}, we have connected two complementary descriptions of the same system: one which treats the wave propagation directly in the strong field, and one which reconstructs the finite-frequency interference pattern from the null-geodesic structure of the spacetime.

For a Schwarzschild lens and the SBI configuration, where the binary angular momentum is oriented along the source--lens axis, we find remarkably good quantitative agreement between the two approaches for sources near the optical axis. In particular, the frequency dependence of the amplification function approaches the same universal curve after the expected caustic rescaling, even for $M\omega\sim\mathcal{O}(1)$. This demonstrates that the null-geodesic arrival-time structure retains predictive power well beyond the strict geometric-optics limit. 

The source emission geometry itself becomes important in this regime. For the SBII configuration, the spin-2, quadrupolar emission introduces an additional direction-dependent phase, shifting the maximum amplification away from perfect source--lens--observer alignment even for a Schwarzschild lens. A phase-corrected RT
calculation reproduces this displacement, although the present implementation does not yet include the full anisotropic emission amplitude or the two independent GW helicities. This configuration is also of particular astrophysical interest for binaries embedded in accretion disks, where gas interactions are expected to drive the binary orbital plane towards alignment or
anti-alignment with the disk plane.

The picture is equally striking for a rotating lens. For SBI near a $a=0.9M$ Kerr BH, the amplification pattern near the caustic is well described by the Schwarzschild RT result once the frame-dragging-induced displacement of the caustic is taken into account. At the frequencies accessible to the
present calculation, the finite astroidal extent of the Kerr caustic is not resolved and the diffraction pattern therefore remains effectively point-caustic-like. This also suggests that extracting the black-hole spin from this feature alone will be challenging in this frequency regime. A systematic study over Kerr spin remains necessary before establishing how general this degeneracy is.

Finally, we have quantified the limitations of replacing the BH by the Newtonian point-lens (PL) model commonly used in GW lensing calculations. The PL is recovered asymptotically in the weak field, but its image positions and arrival-time differences deviate from those of the Schwarzschild lens as the source approaches the MBH. Higher order corrections to the deflection angle can be absorbed into an effective PL mass and substantially improve the fringe spacing. They cannot, however, simultaneously reproduce the strong-field magnifications and arrival-time differences: the effective PL remains a one-parameter description of a system with additional strong-field structure. Deep in the strong field regime, the prescription ceases to have a meaningful geometrical interpretation altogether.

This distinction is relevant for the astrophysical systems motivating this work. At the $\mathcal{O}(10^2$--$10^3)M$ radii associated with plausible accretion disk migration traps, the approach to the point-lens limit is slow, and $\mathcal{O}(1-10)\%$ corrections to the effective lensing scale can remain. Even relatively small  errors in the image arrival times accumulate into appreciable dephasing at high frequency. Consequently, applying a standard PL template to these systems can bias the inferred lens parameters and, in less symmetric
source configurations, can miss structure associated with the quadrupolar emission and polarization of the source entirely.

Together, these results provide a direct link between black-hole scattering and the traditional language of gravitational-wave lensing. They show both why a ray-based diffraction construction can remain useful surprisingly deep into the wave-optics regime and where the standard Newtonian lensing description begins to fail. Extending the RT construction to the complete spin-2 emission pattern, polarization transport, source motion, and a fully Kerr arrival-time surface will allow this correspondence to be tested for realistic waveforms and will be important for identifying triple systems with future gravitational-wave observations.

\begin{acknowledgments}
We are grateful to Vitor Cardoso, Jose Maria Ezquiaga, Nicolas Aimar, Conor Dyson, Giulia Cusin, and Brian Seymour for their helpful comments and suggestions.
This work was supported by Research Grants No.~VIL37766 and No.~VIL53101 from Villum Fonden and the DNRF Chair Program Grant No.~DNRF162 by the Danish National Research Foundation.
This work has received funding from the European Union's Horizon 2020 research and innovation programme under the Marie Sklodowska-Curie Grant Agreement No.~101131233.
J.S.S. acknowledges financial support provided under the European Union’s H2020 ERC Advanced Grant “Black holes: gravitational engines of discovery” grant agreement no. Gravitas–101052587, and also by the H2020-MSCA-2022-SE
project EinsteinWaves, grant agreement no. 101131233.
Views and opinions expressed are however those of the author only and do not necessarily reflect those of the European Union or the European Research Council. Neither the European Union nor the granting authority can be held responsible for them.
J.S.S also thanks the Danish Data Science Academy for support during the development of this work through Visit Grant 2026-6505, as well as the Fundação para a Ciência e Tecnologia (FCT), Portugal, for the financial support to the Center for Astrophysics and Gravitation (CENTRA/IST/ULisboa) through grant No. UID/PRR/00099/2025 and grant No. UID/00099/2025.
\end{acknowledgments}

%

\appendix

\section{Toy model of strong lensing from a quadrupolar source}
\label{ap:toymodelsource}
To our knowledge, all previous investigations of lensing fall into categories where the source can be adequately modeled as an isotropic point source. This happens because the size of the source and the spacial variability of its emission profile are usually negligible in comparison to the cosmological distances separating it from the lens and observer. This is not the case in our strong lensing setup. Modeling the source emission profile becomes of paramount importance in such extreme lensing scenarios; in particular, it explains the shift in the point of maximum amplification seen in the middle panel of Fig.~\ref{fig:F_schwarzschild}. Here we develop a toy model in the geometric optics approximation that clarifies the effects of the quadrupole emission profile. 
\begin{figure}
    \centering
    \usetikzlibrary{arrows.meta,calc,angles,quotes,shapes.geometric,decorations.markings}

\definecolor{myred}{HTML}{C40000}
\definecolor{myblue}{HTML}{0053A1}
\definecolor{mygray}{HTML}{8A8A8A}

\begin{tikzpicture}[
    scale=1.05,
    >=Latex,
    bluepath/.style={
        myblue,
        very thick,
        decoration={
            markings,
            mark=at position 0.45 with {\arrow{Latex}}
        },
        postaction={decorate}
    },
    redsrc/.style={myred, very thick},
    axis/.style={black, thick}
]

\coordinate (S)   at (0,0);
\coordinate (P1)  at (1.10,0.22);
\coordinate (P2)  at (1.10,-0.22);

\coordinate (L)   at (4.10,0);
\coordinate (Obs) at (7.00,1.25);

\draw[axis] (-0.35,0) -- (7.20,0);

\begin{scope}[rotate around={35:(0,0)}]
    \draw[redsrc, smooth, samples=200, variable=\t, domain=0:360]
        plot ({0.54*sin(\t)*cos(\t)}, {0.95*sin(\t)});
\end{scope}

\begin{scope}[rotate around={35:(0,0)}]
    \fill (0,0.28) circle (0.055);
    \fill (0,-0.28) circle (0.055);
\end{scope}

\node[myred] at (0.5, 1.1) {\small $\Psi \sim e^{-i\omega t+2 i \varphi_{\rm S}}$};

\fill[myblue] (P1) circle (0.06);
\fill[myblue] (P2) circle (0.06);

\node[myblue, right] at ($(P1)+(0.,0.25)$) {$1$};
\node[myblue, right] at ($(P2)+(0.,-0.25)$) {$2$};

\draw[bluepath]
    (P1)
    .. controls (3.50,.9) ..
    (Obs);

\draw[bluepath]
    (P2)
    .. controls (4.50,-1.30) ..
    (Obs);

\fill[black] (L) circle (0.42);

\draw[black, very thick, dashed]
    (L) -- (Obs);

\draw[myblue, thick]
    ($(P2)+(0.38,-0.12)$)
    arc[start angle=-55,end angle=70,radius=0.36];

\node[myblue] at (2.05,0.20)
    {\small $\Delta\varphi_{\rm S}$};

\draw[black, thick]
    ($(L)+(0.62,0)$)
    arc[start angle=0,end angle=16,radius=0.95];

\node[black] at (5.25,0.16) {\small $\phi_{\mathrm{O}}-\pi$};

\begin{scope}[shift={($(Obs)+(0.05,0.12)$)}]
    \draw[fill=white] (0,0.10) circle (0.10);

    \draw (0,0.00) -- (0,-0.28);

    \draw (-0.16,-0.10) -- (0,-0.02) -- (0.16,-0.10);

    \draw (0,-0.28) -- (-0.14,-0.46);
    \draw (0,-0.28) -- (0.14,-0.46);
\end{scope}

\end{tikzpicture}
    \caption{Geometric setup in our model for lensing from a quadrupolar source. We consider only the equatorial plane of a Schwarzschild BH (central black disk, the lens) a source os scalar radiation mimicking a binary orbiting in the plane (pair of points at the left), and an observer (person on the right). As in the main text, we fix the source-lens axis, and track the observer's position with the angle $\phi_{\rm O}$. The source is emitting scalar radiation $\Psi$ with a quadrupole profile (red lines): the wavefronts are not circles around the source, instead depending on a local angle $\varphi_{\rm S}$ as shown. The radiation detected by the observer will be, to leading order, the interference of two paths (blue lines), one going over and another going under the lens. Their relative phase will be influenced by the light travel time, but also by their relative emission angle $\delta \varphi_{\rm S}$.}
    \label{fig:paths_diagram}
\end{figure}
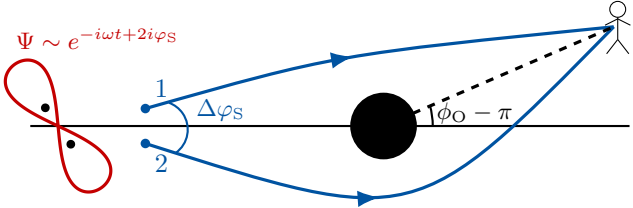

Consider the two dimensional lensing setup illustrated in Fig.~\ref{fig:paths_diagram}: a quadrupolar source of scalar radiation, a lens and  an observer. In a local frame near the source, take the radiation field to be 
\begin{equation}
    \Psi_{\rm S} = \frac{e^{ -i (\omega t - 2 \varphi_{\rm S})}}{R} \, , \label{eq:toy_phase}
\end{equation}
with $(R,\varphi_{\rm S})$ a polar chart centered at the source. This profile mimics gravitational radiation from a binary in the quadrupole approximation (see Eq.~\eqref{eq:unlensed}). There is also a Schwarzschild MBH (center) that acts as a lens and an observer. Let us look at the interference between radiation that reaches the observe following paths 1 and 2: null geodesics going over and under the lens, respectively. For now, we don't care about the exact form of these geodesics, just that there are two and that the travel time is unequal. 

Under these assumptions, the field at the observer $\Psi_{\rm O}$, will be the sum of the field at the observer traveling along the two paths, say $\Psi_{1,2}$, that is
\begin{equation}
    \Psi_{\rm O}( t) = \Psi_{\rm O 1} (t) + \Psi_{\rm O  2} (t), \label{eq:toy_field_obs}
\end{equation}
where, using Eq.~\eqref{eq:toy_phase},
\begin{align}
  \Psi_{{\rm O} j}(t)= & A\, e^{-i \omega (t-t_{j})} e^{-i (\omega t_j - \varphi_{{\rm S}j} )} \, , \quad j=1,2 \,.
\end{align}
where $t-t_j=T_j$ is the travel time along path $j$, and $\varphi_{{\rm S}j}$ is the emission angle of the path $j$. Note that $t_j = t-T_j$ is the emission time, which justifies the use of Eq.~\eqref{eq:toy_phase}. As usual in the geometric optics limit, the amplitude $A$ is a slowly varying function, so it is taken to be the same for both paths. We further choose $A=1$ as we are only interested in the interference of the two paths, not the overall amplitude. Putting everything together yields, up to a global phase
\begin{equation}
\begin{split}
      \Psi_{\rm O}(t) &= \mathcal A \ e^{- i \omega t }\,, \\
    \mathcal{A} &= 1 + e^{i\left( -\omega ( t_1 - t_2) + 2 \Delta \varphi_{\rm S}\right)} \, . 
\end{split}
 \label{eq:toy_calA} 
\end{equation}
$\mathcal{A}=\mathcal{A}(\phi{\rm O})$ measures the constructive or destructive interference between the two paths. It is a function of $\phi_{\rm O}$ as different angles will correspond to different pairs of geodesics. Besides the usual time delay term $t_1 -t_2$, we get a term proportional to $\Delta \varphi_{\rm S} = \varphi_{{\rm S}1}-\varphi_{{\rm S}2}$ (see Fig.~\ref{fig:paths_diagram}). The factor two in front of  the latter reflects the quadrupolar nature of the source. This is the most important formula in this Appendix, as it shows that when $\Delta \varphi_{\rm S} \neq 0 $ there will not be maximum constructive interference when $t_1-t_2 = 0$, as one usual expects. 

We now have to calculate $t_1 - t_2$ and $\Delta \varphi_{\rm S}$. Doing so requires specifying the lens geometry, which take to be Schwarzschild; the plane in Fig.~\ref{fig:paths_diagram} is its equatorial plane. Furthermore, we assume the observer is at a distance $r_{\rm O} \gg r_{\rm S}$. When $\phi_{\rm O} = \pi$, $t_1 = t_2$, the two geodesics are mirrors of each other. These are characterized by their impact parameter $b \equiv b_2 = -b_1 $, which in the weak field approach the well known $b=2\sqrt{M r_{\rm S}}$. Each impact parameter has a corresponding emission angle at the source 
\begin{equation}
    \varphi_{\rm S1}= -\varphi_{\rm S2} = \arcsin{\sqrt{1-2M/r_{\rm S}}b/r_{\rm S}} \, .
\end{equation}
These quantities are determined by solving the geodesic equation in Schwarzschild from infinity to a point at coordinate distance $r_{\rm S}$ of the lens on the optical axis~\cite{Chandrasekhar:1985kt}. Now we want to take the leading change in the travel time $t_1 - t_2$ due to changing $\phi_{\rm O}$. To do so, we use the relation between the travel time along a geodesic $T_j$ and the accumulated azimuthal phase $\Phi_j$ valid for null geodesics $dT_j / d\Phi_j = b_j$. In the symmetric setup we described here, we get
\begin{equation}
t_1 - t_2 = T_2-T_1 = 2 b (\phi_{\rm O}-\pi) + \mathcal O((\phi_{\rm O}-\pi)^2) \, . \label{eq:toy_time_delay}
\end{equation}
%



%
\begin{figure}
    \centering
    \includegraphics[width=\linewidth]{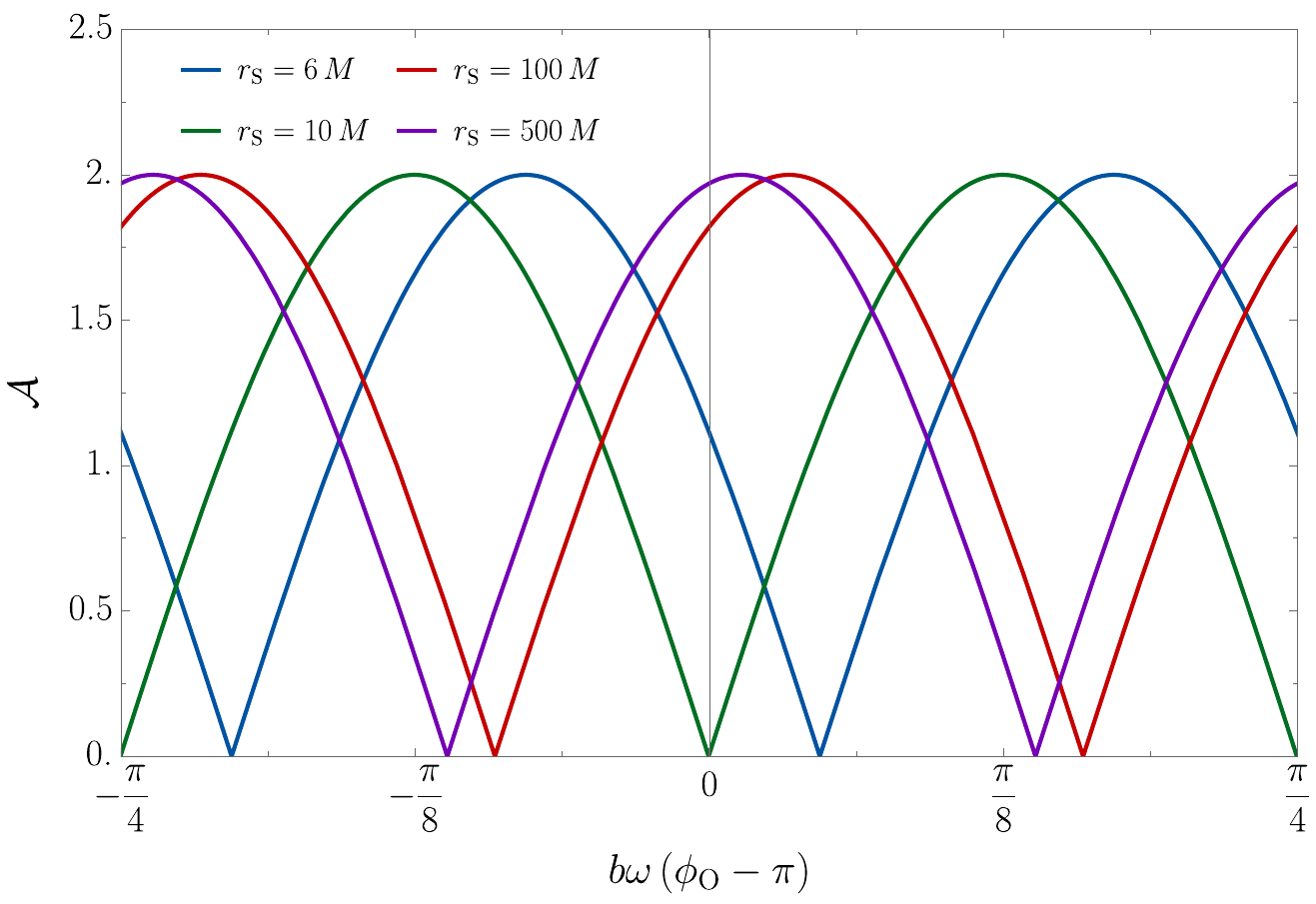}
    \caption{Amplitude of the field at the observer in the toy model (Eq.~\eqref{eq:toy_calA}) as a function of the rescaled observer angular position $b \omega \phi_{\rm O}$ (see Eq.~\eqref{eq:toy_time_delay}). Different curves correspond to different source distances from the lens $r_{\rm S}$ in lens mass $M$ units (the observer is on the celestial sphere). The quadrupole nature of the source means that the amplitude is, in general, not maximized for $\rm \phi_{\rm O}=\pi$. This is the case only when $r_{\rm S}\to \infty$.
    }
    \label{fig:toy_results}
\end{figure}

The results obtained with the toy model are presented in Fig.~\ref{fig:toy_results}, for different source positions $r_{\rm S}$. Indeed, comparing with the numerical results for Schwarzschild in Fig.~\ref{fig:F_schwarzschild} we find that our toy model is able to capture the deviation of the peak from alignment. The model also accurately predicts the direction of the shift: as $r_S$ is decreased, maximum amplification shifts to larger positive $\phi_{\rm O}$. When $r_{S}\to \infty$, we get that $\Delta \varphi_{\rm S} \to 0$, and so there is maximum amplification for $\phi_{\rm O} = \pi$\footnote{We did not include the results showing that this is the direction of the shift, but we verified it is the case.}. 

Of course, a lot of simplifying assumptions go into this model, so we don't expect it to fully reproduce the numeric results. Still, it serves the important purpose of illustrating how, in general, a non isotropic source of radiation does not produce a symmetric wave pattern even if the source-lens-observer geometry exhibits a particular symmetry. 

A pressing follow up question is then: why does the SB I setup in Fig.~\ref{fig:binary_geometry} produce the symmetric pattern in the top panel of Fig.~\ref{fig:Mollweides}? The key is that GWs are tensor fields, not scalars as we treated the here. Indeed, in a local frame around the binary, each GW polarization is still emitted with the phase shift shown in Eq.~\eqref{eq:toy_phase}. However, the local polarization frame is different for different points in a sphere around the binary (they are defined with respect to a local basis of tangent space, spanned by $\partial_{\vartheta_{\rm S}}, \partial_{\varphi_{\rm S}}$ as defined in Eq.~\eqref{eq:unlensed}). What we want is to compare the polarizations at the observer. For the SB I geometry, these are related by a phase shift of $e^{-2 i \varphi_{\rm S}}$, which cancels the $\varphi$ dependent factor in Eq.~\eqref{eq:toy_phase}, effectively recovering an isotropic source. Contrarily in the SB II setup, reflection symmetry requires that the radiation is linearly polarized in the equatorial plane, so the radiation does interfere the same as a scalar, as assumed in this toy model.

\section{Kerr Caustics}
\label{ap:kerrcaust}

Here we provide details for the angular offset of the Kerr caustics $\phi_{\rm c}$ used in Sec.~\ref{sec:results_Kerr}. An approximate formula for the angular coordinates of the geometric optics caustic was obtained in~\cite{Sereno:2007gd}. In our setup, with $r_{\rm O} \gg r_{\rm S}$ and for an equatorial source and observer, the caustic is crossed when the observer is at\footnote{The caustic curve is defined on the source plane, so this is usually written as $\phi_{\rm S} = \phi_{\rm c}$. Here we are instead choosing to fix the source angular coordinates, and vary those of the observer. This means we will see the caustic crossing when $\phi_{\rm O}= \phi_{\rm c}$.}
\begin{align}
	\phi_{\rm O} =
     \pi &+ \frac{ a}{M} \left( 4 \varepsilon^2 + \frac{5 \pi}{4} \varepsilon^3 -\left(\frac{225 \pi^2}{128}-16 \right)  \varepsilon^4 \right)  \label{eq:caustic_formula}\\
    & \hspace{9 em} \mp \left(\frac{a}{M}\right)^2 \frac{15 \pi}{16} \varepsilon^4 + \mathcal{O}(\varepsilon^5) \, , \nonumber 
 \end{align}
for $\varepsilon = \sqrt{M/4 r_{\rm S}}$. When $a\to 0$, the caustic crossing happens at the antipodal point, as required. For the fiducial setup defined in Tab.~\ref{tab:parameters_fiducial}, $\varepsilon\approx0.16$, so the neglected terms are indeed very small. 

The angular size of the caustic is given by the term in the second line of Eq.~\eqref{eq:caustic_formula}, and for the fiducial system is $\Delta \phi_{\rm S} \sim 10^{-3} (a/M)^2 $. This means, for the WO scenarios we are investigating, it is impossible to resolve the size of the caustic, as $\Delta \phi_{\rm O}\ll M\omega$. Thus, we expect to get maximum amplification when the observer is at $\phi_{\rm O} \approx \phi_{\rm c}$, where 
\begin{equation}
	\phi_{\rm c} = 
     \pi + \frac{ a}{M} \left( 4 \varepsilon^2 + \frac{5 \pi}{4} \varepsilon^3 -\left(\frac{225 \pi^2}{128}-16 \right)  \varepsilon^4 \right) \, .\label{eq:caustic}
\end{equation}
\end{document}